\documentclass[preprint,showpacs,preprintnumbers,amsmath,amssymb,superscriptaddress]{revtex4}

\usepackage{graphicx}% Include figure files
\usepackage{dcolumn}% Align table columns on decimal point
\usepackage{bm}% bold math
\begin{document}

%\preprint{CREAM/MS-2004-01-PRA-GH-01}

\title{Canonical equations of Hamilton with beautiful symmetry}% Force line breaks with \\

\author{Guo Liang}
\affiliation{Laboratory of Nanophotonic Functional Materials and Devices, South
China Normal University, Guangzhou 510631, P. R. China}
\author{Qi
Guo} \email{guoq@scnu.edu.cn} \affiliation{Laboratory of Nanophotonic Functional Materials and Devices, South
China Normal University, Guangzhou 510631, P. R. China}

\date{\today}% It is always \today, today,

\begin{abstract}
The Hamiltonian formulation plays the essential role in constructing the framework of modern physics. In this paper,
a new form of canonical equations of Hamilton with the complete symmetry is obtained, which are valid not only
for the first-order differential system, but also for the second-order differential
system. The conventional form of the canonical equations without the symmetry [Goldstein \emph{et al}., Classical Mechanics, 3rd ed, Addison-Wesley, 2001] are only for the second-order differential
system. It is pointed out for the first time that the number of the canonical equations for the first-order differential
system is half of that
for the second-order differential system. The nonlinear Schr\"{o}dinger equation,
 a universal first-order differential system, can be expressed with the new canonical equations in a consistent way.
\end{abstract}

.

%\keywords{Use showkeys class option if keyword display desired}%
\maketitle
%\section{Introduction}
The Hamiltonian viewpoint provides a framework for theoretical
extensions in many areas of
physics.
In classical mechanics it forms the basis for further developments,
such as Hamilton-Jacobi theory, perturbation approaches and chaos~\cite{Goldstein-book-05}. The canonical equations of
Hamilton in classical mechanics are expressed as~\cite{Goldstein-book-05}
\begin{eqnarray}
\dot{q_i}&=&\frac{\partial H}{\partial p_i},\quad (i=1,\cdots,n),\label{Hamilton formulations 1 for discrete system}\\
 -\dot{p_i}&=&\frac{\partial H}{\partial q_i},\quad (i=1,\cdots,n),\label{Hamilton formulations 2 for discrete system}
\end{eqnarray}
where $q_i$ and $p_i$ are said to be the generalized coordinate and the generalized momentum, $\dot{q}_i=d q_i/dt$ is the generalized velocity, $\dot{p}_i=d p_i/dt$, and $H$ is the Hamiltonian of the system. The generalized momentum $p_i$ is defined as
$p_i=\frac{\partial L}{\partial\dot{q}_i}$, where $L$ is the Lagrangian of the system. And the Hamiltonian $H$ is obtained by the Legendre transformation
$H=\sum_{i=1}^n\dot{q}_ip_i-L$.
 A set of $(q_i,p_i)$ forms a $2n$-dimensional phase space.
%The Hamiltonian formulation (\ref{derivative of q}) and (\ref{derivative of pi}) are advantageous
% to us because they afford us the deeper insight into the formal structure of mechanics but not
%only a calculational tool.

The canonical equations (\ref{Hamilton formulations 1 for discrete system}) and (\ref{Hamilton formulations 2 for discrete system}) are not only valid for the discrete system, but also can be extended to the continuous system as~\cite{Goldstein-book-05}
\begin{eqnarray}
\dot{q}_s&=&\frac{\delta h}{\delta p_s},\quad (s=1,\cdots,N),\label{derivative of q for continuous system}\\
-\dot{p}_s&=&\frac{\delta h}{\delta q_s},\quad (s=1,\cdots,N),\label{derivative of
pi for continuous system}
\end{eqnarray}
where the subscript $s$ represents the components of the quantity of the continuous system~\cite{Goldstein-book-05}, hereafter it all denotes $1,\cdots,N$, $\frac{\delta h}{\delta q_s}=\frac{\partial h}{\partial q_s}-\frac{\partial}{\partial x}\frac{\partial h}{\partial q_{s,x}}$ and $\frac{\delta h}{\delta p_s}=\frac{\partial h}{\partial p_s}-\frac{\partial}{\partial x}\frac{\partial h}{\partial p_{s,x}}$ denote the functional derivatives of $h$ with respect to $q_s$ and $p_s$ with $q_{s,x}=\frac{\partial q_s}{\partial x}$ and $p_{s,x}=\frac{\partial p_s}{\partial x}$, $q_s$ and $p_s$ are the generalized coordinate and the generalized momentum, respectively, and $h$ is the Hamiltonian density of the continuous system. Like the discrete system, the generalized momentum $p_s$ for the continuous system is defined as
 \begin{equation}\label{conjugate momenta}
 p_s=\frac{\partial l}{\partial\dot{q}_s},
 \end{equation}
 and the Hamiltonian density $h$ for the continuous system is obtained by the Legendre transformation as
 \begin{equation}\label{legendre transform}
 h=\sum_{s=1}^N\dot{q}_sp_s-l,
 \end{equation}
where $l$ is the Lagrangian density.
But it is significantly different for the continuous system that $q_s$ and $p_s$ are now not only functions of time $t$, but also the spatial coordinate $x$, where the spatial coordinate $x$
is not the generalized coordinate, but only serves as the continuous index replacing the discrete $i$
in Eqs.(\ref{Hamilton formulations 1 for discrete system}) and (\ref{Hamilton formulations 2 for discrete system}). To distinguish $t$ from the coordinate $x$, we refer time $t$
as the evolution coordinate. $p_s$ and $q_s$ define the infinite-dimensional phase space.
$h$ is a function of $q_s,p_s$ and $q_{s,x}$ but not $p_{s,x}$, so $\frac{\delta h}{\delta p_s}=\frac{\partial h}{\partial p_s}$, then Eq.(\ref{derivative of q for continuous system}) can be also expressed as
\begin{equation}\label{another form}
\dot{q}_s=\frac{\partial h}{\partial p_s}.
\end{equation}

%$N$ in Eqs.(\ref{derivative of q for continuous system}) and (\ref{derivative of pi for continuous system}) denotes the number of the canonical variables of the continuous system (it is clearer in the following example). It has the different meaning from $n$ in Eqs.(\ref{Hamilton formulations 1 for discrete system}) and (\ref{Hamilton formulations 2 for discrete system}) (see pages 13-15 and pages 583-585 of \cite{Goldstein-book-05} for details.), which represents the degrees of freedom of the discrete system. To stress the difference, we use the subscript $s$ instead of $i$ in Eqs.(\ref{derivative of q for continuous system}) and (\ref{derivative of pi for continuous system}).

``The advantages of the Hamiltonian formulation lie not
in its use as a calculational tool," as has been pointed out in ref.\cite{Goldstein-book-05}, ``but rather in the deeper insight it affords into the formal structure of mechanics."
% The equal status accorded to coordinates and momenta as independent variables encourages a greater freedom in selecting the physical quantities to designated as `coordinates' and `momenta'.
We have a greater freedom to select the physical quantities to designate as ``coordinates'' and ``momenta'', which have equal status as independent variables.
  As a result we can present the physical content of mechanics in a newer and more abstract way. The more abstract formulation is primarily of interest to us today because of its essential role in constructing the framework of modern physics, such as statistical mechanics and quantum theory.

%There are many examples of the application of the Hamiltonian formulation. The first law of thermodynamics was interpreted from
%the Hamiltonian viewpoint setting up a bridge between the classical
%thermodynamics and the particle mechanics~\cite{menon-EJP}. In
%electromagnetics, the Hamiltonian formulation was used to describe
%the dynamics of LC-circuits and the behaviour of a large class of
%switching power converters~\cite{Maschke-IEEE,Escobar-Automatica}.
%The Hamiltonian formulation can also be applied in the field of
%optics, the main virtue of which lies in its ability directly to
%yield general statements about the over-all properties of optical
%systems, without any need to inquire into the details of their
%constructions~\cite{Buchdahl-book-93}.
\subsection*{Canonical equations of Hamilton with symmetry}
The Newton's second law of motion in classical mechanics, based on which the Hamiltonian formulation is established, is the second-order differential equation about the evolution coordinate (here the evolution coordinate is time), which is in this paper referred as the second-order differential system. Similarly, the first-order differential system is the first-order differential equation about the evolution coordinate. The Lagrangian density of the second-order differential system of the continuous systems is expressed in general as ~\cite{Goldstein-book-05}
\begin{equation}\label{Lagrangian density of 2nd system}
l=\sum_{s=1}^N\sum_{k=1}^NA_{sk}\dot{q}_s\dot{q}_k+\sum_{s=1}^NB_s\dot{q}_s+C,
\end{equation}
%$A_{sk}=A_{sk}(q_1,\cdots,q_N),B_{s}=B_{s}(q_1,\cdots,q_N),C=C(q_1,\cdots,q_N)$
where $A_{sk},B_{s},C$ depend on not only $q_s$ but also $q_{s,x}$ in general.
The generalized momentum can be obtained by the definition (\ref{conjugate momenta}) as
\begin{equation}
p_s=\sum_{k=1}^N(A_{sk}+A_{ks})\dot{q}_k+B_s\label{conjugate momenta 2},
\end{equation}
which is a function of $q_s$, $\dot{q}_s$ and $q_{s,x}$.
The number of Eqs. (\ref{conjugate momenta 2}) is $N$, and there are $4N$ variables, which are $q_s$, $\dot{q}_s$, $p_s$ and $q_{s,x}$. So the degree of freedom of  Eqs. (\ref{conjugate momenta 2}) is $3N$. Then $q_s,p_s$ and $q_{s,x}$ are taken as independent variables. The generalized velocities $\dot{q}_s$ can be expressed with these independent variables.

Besides the second-order differential systems, there are a number of the first-order differential systems to model physical phenomena, for example, the nonlinear Schr\"{o}dinger equation (NLSE)
\begin{equation}\label{NNLSE}
i\frac{\partial \varphi}{\partial
t}+\frac{1}{2}\frac{\partial^2\varphi}{\partial x^2}+|\varphi|^2\varphi=0,
\end{equation}
which is
a universal
model that describes many nonlinear physical systems and can be
applied to hydrodynamics~\cite{Nore-PhysicaD-93}, nonlinear
optics~\cite{Haus-book,Hasegawa-book-95,Agrawal-book}, nonlinear
acoustics~\cite{Bisyarin-AIPConf.Proc.-08}, Bose-Einstein
condensates~\cite{seaman-pra-05}, and so on. In nonlinear optics~\cite{Hasegawa-book-95,Haus-book,Agrawal-book}, the NLSE (\ref{NNLSE}) governs the propagation of the slowly-varying light-envelope,
where the evolution coordinate $t$ is the propagation
direction coordinate, the light-envelope $\varphi$ is a cw paraxial beam
in a planar waveguide~\cite{Haus-book} or a narrow spectral-width pulse
in optical fibers~\cite{Hasegawa-book-95,Agrawal-book}, and $x$ is a transverse space coordinate for the beam and a frame moving at the group velocity (the so-called retarded frame)
for the pulse, respectively. We will show in the paper that the canonical equations for the first-order differential system are significantly different from Eqs.(\ref{derivative of q for continuous system}) and (\ref{derivative of pi for continuous system}), which are the canonical equations only for the second-order differential system. But in ref.\cite{Goldstein-book-05}, the canonical equations (\ref{derivative of q for continuous system}) and (\ref{derivative of pi for continuous system}) were obtained regardless whether the continuous system is the first-order differential system or the second-order differential system.
In ref.\cite{Hasegawa-book-95}, the canonical equations
for the NLSE were considered to be the same as Eqs.(\ref{derivative of q for continuous system}) and (\ref{derivative of pi for continuous system}). But obviously the canonical equations (\ref{derivative of q for continuous system}) and (\ref{derivative of pi for continuous system}) are not valid for the NLSE, from which it is impossible to obtain the NLSE or its complex-conjugate equation, as will be shown in the following. Attempt was made to deal with the difficulty in ref.\cite{Dauxois-book-06}, but the canonical equations for the NLSE they obtained were inconsistent, as will be shown in the following. In this paper
 we obtain the canonical equations with the complete symmetry, which are valid not only
for the first-order differential system, but also for the second-order differential
system.

For the first-order differential system of the continuous systems,
the Lagrangian density must be the linear function of the generalized velocities $\dot{q}_s$.
 If the Lagrangian density is a quadratic function of the generalized velocities like Eq.(\ref{Lagrangian density of 2nd system}), the
equation of motion, i.e., the Euler-Lagrange equation
 \begin{equation}\label{Euler-Lagrange equation}
\frac{\partial}{\partial t}\frac{\partial l}{\partial\dot{q}_s}-\frac{\delta l}{\delta q_s}=0
\end{equation}
will be the second-order differential equation about the evolution coordinate $t$, which is
 in contradiction with the definition of the first-order differential system. Therefore, the Lagrangian density of the first-order differential system can only be expressed as
\begin{equation}\label{Lagrangian density for the fist-order system}
l=\sum_{s=1}^NR_s(q_s)\dot{q}_s+Q(q_s,q_{s,x}).
\end{equation}
%substitution of which into the Euler-Lagrange equation $\frac{d}{d t}\frac{\partial l}{\partial\dot{q}_s}-\frac{\delta l}{\delta q_s}=0$ yields the first-order differential equation about the evolution equation $t$.
Besides, $R_s$ in Eq.(\ref{Lagrangian density for the fist-order system}) is not the function of a set of $q_{s,x}$. If $R_s$ is also the function of a set of $q_{s,x}$, there will be such terms as $q_{s,x}\dot{q}_s$ appearing in Eq.(\ref{Lagrangian density for the fist-order system}). Substitution of Eq.(\ref{Lagrangian density for the fist-order system}) into Eq.(\ref{Euler-Lagrange equation}) leads to the appearance of the mixed partial derivative term $\frac{\partial^2 q_s}{\partial x\partial t}$,
the term can be changed to $\frac{\partial^2 q_s}{\partial t^2}$ by the rotation of the coordinate frame. After the coordinate transformation, the Euler-Lagrange equation (\ref{Euler-Lagrange equation}) is changed to the second-order partial differential equation of the standard form. Then the system expressed with the standard form is a second-order differential system. Consequently, the generalized momentum $p_s$, which is obtained by the definition (\ref{conjugate momenta}) as
\begin{equation}\label{equations of p q}
p_s=R_s(q_s),
\end{equation}
is only a function of $q_s$. This is of significant difference from the case of the second-order differential system, where the generalized momentum $p_s$ is the function of not only $q_s$, but also $\dot{q}_s$ and $q_{s,x}$, as shown in Eq.(\ref{conjugate momenta 2}). There are $2N$ variables, $q_s$ and $p_s$, in Eqs. (\ref{equations of p q}). And the number of Eqs. (\ref{equations of p q}) is $N$, which also means there exist $N$ constraints between $q_s$ and $p_s$. So the degree of freedom of the system given by Eqs. (\ref{equations of p q}) is $N$. Without loss of generality, we take $q_1,\cdots,q_\nu$ and $p_1,\cdots,p_\mu$ as the independent variables, where $\nu+\mu=N$. The remaining generalized coordinates and generalized momenta can be expressed with these independent variables as
$
q_\alpha=f_\alpha(q_1,\cdots,q_\nu,p_1,\cdots,p_\mu)(\alpha=\nu+1,\cdots,N),$ and $
p_\beta=g_\beta(q_1,\cdots,q_\nu,p_1,\cdots,p_\mu) (\beta=\mu+1,\cdots,N).
$

We now derive the canonical equations for the first-order differential system.
 The total differential of the Hamiltonian density $h$ can be obtained by
using Eq.(\ref{legendre transform})
{\setlength\arraycolsep{0pt}
\begin{eqnarray}
d h&=&\sum_{s=1}^Np_sd \dot{q}_s+\sum_{\eta=1}^{\mu}\dot{q}_\eta d p_\eta\nonumber\\
&+&\sum_{\beta=\mu+1}^{N}\dot{q}_\beta\left(\sum_{\lambda=1}^{\nu}\frac{\partial g_\beta}{\partial q_\lambda}d q_\lambda+\sum_{\eta=1}^{\mu}\frac{\partial g_\beta}{\partial p_\eta}d p_\eta\right)
-d l,\label{eq1}
\end{eqnarray}}
while the total differential of the Lagrangian density $l(q_s,\dot{q}_s,q_{s,x})$ with respect to its arguments is
{\setlength\arraycolsep{0pt}
\begin{eqnarray}
d l&=&\sum_{\lambda=1}^{\nu}\frac{\partial l}{\partial q_\lambda}d q_\lambda+\sum_{\alpha=\nu+1}^{N}\frac{\partial l}{\partial q_\alpha}\left(\sum_{\lambda=1}^{\nu}\frac{\partial f_\alpha}{\partial q_\lambda}d q_\lambda+\sum_{\eta=1}^{\mu}\frac{\partial f_\alpha}{\partial p_\eta}d p_\eta\right)\nonumber\\
&&+\sum_{s=1}^N\frac{\partial l}{\partial\dot{q}_s}d \dot{q}_s+\sum_{s=1}^N\frac{\partial l}{\partial q_{s,x}}d q_{s,x}.\label{eq2}
\end{eqnarray}}
%To obtain the second equality of Eq.(\ref{eq2}), we have used Eq.(\ref{conjugate momenta}) and the Euler-Lagrange equations
% \begin{equation}\label{Euler-Lagrange equation}\frac{d}{dt}\frac{\partial l}{\partial\dot{q}_s}-\frac{\delta l}{\delta q_s}=0.
% \end{equation}
 Substitution of Eq.(\ref{eq2}) into Eq.(\ref{eq1}) yields
 {\setlength\arraycolsep{0pt}
\begin{eqnarray}
d h&=&\sum_{\lambda=1}^{\nu}\left(\sum_{\beta=\mu+1}^{N}\dot{q}_\beta\frac{\partial g_\beta}{\partial q_\lambda}-\sum_{\alpha=\nu+1}^{N}\frac{\partial l}{\partial q_\alpha}\frac{\partial f_\alpha}{\partial q_\lambda}-\frac{\partial l}{\partial q_\lambda}\right)d q_\lambda\nonumber\\
&&+\sum_{\eta=1}^{\mu}\left(\dot{q}_\eta+\sum_{\beta=\mu+1}^{N}\dot{q}_\beta\frac{\partial g_\beta}{\partial p_\eta}-\sum_{\alpha=\nu+1}^{N}\frac{\partial l}{\partial q_\alpha}\frac{\partial f_\alpha}{\partial p_\eta}\right)d p_\eta\nonumber\\
&&-\sum_{s=1}^N\frac{\partial l}{\partial q_{s,x}}d q_{s,x}\label{differential equations of H1}
\end{eqnarray}}
Since the total differential of $h(q_1,\cdots,q_\nu,p_1,\cdots,p_\mu,q_{s,x})$ with respect to its arguments can be written as
$
d h=\sum_{\lambda=1}^{\nu}\frac{\partial h}{\partial q_\lambda}d q_\lambda+\sum_{\eta=1}^{\mu}\frac{\partial h}{\partial p_\eta}d p_\eta+\sum_{s=1}^N\frac{\partial h}{\partial q_{s,x}}d q_{s,x},
$
by comparing this equation with Eq.(\ref{differential equations of H1}), we obtain $2N$ equations
%{\setlength\arraycolsep{0pt}
%\begin{eqnarray}
%\frac{\delta h}{\delta q_\lambda}&=&-\dot{p}_\lambda+\sum_{\beta=\mu+1}^{N}\dot{q}_\alpha\frac{\partial g_\beta}{\partial q_\lambda}-\sum_{\alpha=\nu+1}^{N}\dot{p}_\beta\frac{\partial f_\alpha}{\partial q_\lambda},\label{canonical equation1}\\
%\frac{\delta h}{\delta p_\eta}&=&\dot{q}_\eta+\sum_{\beta=\mu+1}^{N}\dot{q}_\alpha\frac{\partial g_\beta}{\partial p_\eta}-\sum_{\alpha=\nu+1}^{N}\dot{p}_\beta\frac{\partial f_\alpha}{\partial p_\eta},\label{canonical equation2}
%\end{eqnarray}
%}
{\setlength\arraycolsep{0pt}
\begin{eqnarray}
\frac{\partial h}{\partial q_\lambda}&=&\sum_{\beta=\mu+1}^{N}\dot{q}_\beta\frac{\partial g_\beta}{\partial q_\lambda}-\sum_{\alpha=\nu+1}^{N}\frac{\partial l}{\partial q_\alpha}\frac{\partial f_\alpha}{\partial q_\lambda}-\frac{\partial l}{\partial q_\lambda},\label{h q}\\
\frac{\partial h}{\partial p_\eta}&=&\dot{q}_\eta+\sum_{\beta=\mu+1}^{N}\dot{q}_\beta\frac{\partial g_\beta}{\partial p_\eta}-\sum_{\alpha=\nu+1}^{N}\frac{\partial l}{\partial q_\alpha}\frac{\partial f_\alpha}{\partial p_\eta},\label{h p}\\
\frac{\partial h}{\partial q_{s,x}}&=&-\frac{\partial l}{\partial q_{s,x}},\label{h qx}
\end{eqnarray}}
where $\lambda=1,\cdots,\nu,\eta=1,\cdots,\mu,s=1,\cdots,N$. From the Euler-Lagrange equations (\ref{Euler-Lagrange equation}),
we obtain
\begin{equation}\label{Euler-Lagrange equation1}
\frac{\partial l}{\partial q_s}=\frac{\partial}{\partial t}\frac{\partial l}{\partial\dot{q}_s}+\frac{\partial}{\partial x}\frac{\partial l}{\partial q_{s,x}}.
\end{equation}
Substituting Eq.(\ref{Euler-Lagrange equation1}) into Eqs.(\ref{h q}) and (\ref{h p}), we have
{\setlength\arraycolsep{0pt}
\begin{eqnarray}
\frac{\partial h}{\partial q_\lambda}&=&-\dot{p}_\lambda+\sum_{\beta=\mu+1}^{N}\dot{q}_\beta\frac{\partial g_\beta}{\partial q_\lambda}-\sum_{\alpha=\nu+1}^{N}\dot{p}_\alpha\frac{\partial f_\alpha}{\partial q_\lambda}\nonumber\\
&&-\sum_{\alpha=\nu+1}^{N}\frac{\partial}{\partial x}\frac{\partial l}{\partial q_{\alpha,x}}\frac{\partial f_\alpha}{\partial q_\lambda}-\frac{\partial}{\partial x}\frac{\partial l}{\partial q_{\lambda,x}},\label{h q-2}\\
\frac{\partial h}{\partial p_\eta}&=&\dot{q}_\eta+\sum_{\beta=\mu+1}^{N}\dot{q}_\beta\frac{\partial g_\beta}{\partial p_\eta}-\sum_{\alpha=\nu+1}^{N}\dot{p}_\alpha\frac{\partial f_\alpha}{\partial p_\eta}\nonumber\\
&&-\sum_{\alpha=\nu+1}^{N}\frac{\partial}{\partial x}\frac{\partial l}{\partial q_{\alpha,x}}\frac{\partial f_\alpha}{\partial p_\eta},\label{h p-2}
\end{eqnarray}}
Then substituting Eq.(\ref{h qx}) into Eqs.(\ref{h q-2}) and (\ref{h p-2}), we obtain the $N$ canonical equations for the first-order differential system
{\setlength\arraycolsep{0pt}
\begin{eqnarray}
\frac{\delta h}{\delta q_\lambda}&=&-\dot{p}_\lambda+\sum_{\beta=\mu+1}^{N}\dot{q}_\beta\frac{\partial g_\beta}{\partial q_\lambda}-\sum_{\alpha=\nu+1}^{N}\dot{p}_\alpha\frac{\partial f_\alpha}{\partial q_\lambda}\nonumber\\
&&+\sum_{\alpha=\nu+1}^{N}\frac{\partial}{\partial x}\frac{\partial h}{\partial q_{\alpha,x}}\frac{\partial f_\alpha}{\partial q_\lambda},\label{h q-last}\\
\frac{\delta h}{\delta p_\eta}&=&\dot{q}_\eta+\sum_{\beta=\mu+1}^{N}\dot{q}_\beta\frac{\partial g_\beta}{\partial p_\eta}-\sum_{\alpha=\nu+1}^{N}\dot{p}_\alpha\frac{\partial f_\alpha}{\partial p_\eta}\nonumber\\
&&+\sum_{\alpha=\nu+1}^{N}\frac{\partial}{\partial x}\frac{\partial h}{\partial q_{\alpha,x}}\frac{\partial f_\alpha}{\partial p_\eta}.\label{h p-last}
\end{eqnarray}}
To obtain Eq.(\ref{h p-last}), we have used $\frac{\delta h}{\delta p_\eta}=\frac{\partial h}{\partial p_\eta}$, because $h$ is not a function of $p_{\eta,x}$.
The canonical equations (\ref{h q-last}) and (\ref{h p-last}) can be expressed in a completely symmetric form as
{\setlength\arraycolsep{0pt}
\begin{eqnarray}
\frac{\delta h}{\delta q_\lambda}&=&\sum_{s=1}^N\left(\dot{q}_s\frac{\partial p_s}{\partial q_\lambda}-\dot{p}_s\frac{\partial q_s}{\partial q_\lambda}\right)+\sum_{\alpha=\nu+1}^{N}\frac{\partial}{\partial x}\frac{\partial h}{\partial q_{\alpha,x}}\frac{\partial f_\alpha}{\partial q_\lambda},\label{canonical equations 1 for 1st order }\\
\frac{\delta h}{\delta p_\eta}&=&\sum_{s=1}^N\left(\dot{q}_s\frac{\partial p_s}{\partial p_\eta}-\dot{p}_s\frac{\partial q_s}{\partial p_\eta}\right)+\sum_{\alpha=\nu+1}^{N}\frac{\partial}{\partial x}\frac{\partial h}{\partial q_{\alpha,x}}\frac{\partial f_\alpha}{\partial p_\eta}\label{canonical equations 2 for 1st order }
\end{eqnarray}
}($\lambda=1,\cdots,\nu$, $\eta=1,\cdots,\mu$, and $\nu+\mu=N$),
because $\dot{p}_\lambda=\sum_{\lambda'=1}^\nu\dot{p}_{\lambda'}\frac{\partial q_{\lambda'}}{\partial q_\lambda},  \dot{q}_\eta=\sum_{\eta'=1}^\mu\dot{q}_{\eta'}\frac{\partial p_{\eta'}}{\partial p_\eta}$, $\sum_{\eta=1}^{\mu}\dot{q}_\eta\frac{\partial p_\eta}{\partial q_\lambda}=0,$ and $\sum_{\lambda=1}^{\nu}\dot{p}_\lambda\frac{\partial q_\lambda}{\partial p_\eta}=0$.
The canonical equations (\ref{canonical equations 1 for 1st order }) and (\ref{canonical equations 2 for 1st order }) can be easily extended to the discrete system, which can be expressed as
{\setlength\arraycolsep{0pt}
\begin{eqnarray}
\frac{\partial h}{\partial q_\lambda}&=&\sum_{s=1}^N\left(\dot{q}_s\frac{\partial p_s}{\partial q_\lambda}-\dot{p}_s\frac{\partial q_s}{\partial q_\lambda}\right),\label{canonical equations 1 for 1st order for discrete system }\\
\frac{\partial h}{\partial p_\eta}&=&\sum_{s=1}^N\left(\dot{q}_s\frac{\partial p_s}{\partial p_\eta}-\dot{p}_s\frac{\partial q_s}{\partial p_\eta}\right),\label{canonical equations 2 for 1st order for discrete system}
\end{eqnarray}
}
where $\lambda=1,\cdots,\nu$, $\eta=1,\cdots,\mu$, and $\nu+\mu=N$.

Although they are derived from the first-order differential system,
the canonical equations of Hamilton (\ref{canonical equations 1 for 1st order })
and (\ref{canonical equations 2 for 1st order }) with symmetry can be proved to be reduced to Eqs.(\ref{derivative of q for continuous system})
and (\ref{derivative of pi for continuous system}) without symmetry, if all the $N$ generalized coordinates $q_s$ and $N$ generalized momenta $p_s$ in Eqs.(\ref{canonical equations 1 for 1st order })
and (\ref{canonical equations 2 for 1st order }) are independent, because $\frac{\partial g_\beta}{\partial q_\lambda}=
\frac{\partial f_\alpha}{\partial q_\lambda}=\frac{\partial g_\beta}{\partial p_\eta}=
\frac{\partial f_\alpha}{\partial p_\eta}=0$ ($\alpha=\nu+1,\cdots,N,\beta=\mu+1,\cdots,N$) in this case.
%And the number of the canonical equations will be increased by $N$, because the total number of the independent generalized coordinates and generalized momenta is also increased by $N$.
 In fact, this is just the case of the second-order differential system. So the canonical equations (\ref{canonical equations 1 for 1st order })
and (\ref{canonical equations 2 for 1st order }) with the completely symmetry are valid not only
for the first-order differential system, but also for the second-order differential
system. The conventional form of the canonical equations (\ref{derivative of q for continuous system}) and (\ref{derivative of pi for continuous system}) without the symmetry are only for the second-order differential
system. In addition, the number of Eqs. (\ref{canonical equations 1 for 1st order })
and (\ref{canonical equations 2 for 1st order }) is $N$ in the case for the first-order differential system, and is half of that of Eqs.(\ref{derivative of q for continuous system}) and (\ref{derivative of pi for continuous system}).
%where all the generalized coordinates and the generalized momenta are independent.
%In fact, the number of the canonical equations depends on the degrees of freedom of the system. For the second-order differential system described by Eq.(\ref{Lagrangian density of 2nd system}), the degrees of freedom is $2N$, as can be seen from Eqs. (\ref{conjugate momenta 2})~\cite{freedom-2nd-system}. And the degrees of freedom of the first-order differential system described by Eq.(\ref{Lagrangian density for the fist-order system}) is $N$, as can be seen from Eq.(\ref{equations of p q}).
This can be explained in the following way. For the second-order differential system, the Euler-Lagrange equations (\ref{Euler-Lagrange equation})
 are the second-order differential equations about the evolution coordinate, where the Lagrangian density is expressed as Eq.(\ref{Lagrangian density of 2nd system}). The number of Eqs.(\ref{Euler-Lagrange equation}) is $N$. It is well known that one second-order differential equation can be reduced to two first-order differential equations~\cite{M.Lea-book-2004}. Then $2N$ canonical equations, which are the first-order differential equations about the evolution coordinate, can be obtained from the $N$ Euler-Lagrange equations.  But it is significantly different for the first-order differential system that the Euler-Lagrange equations (\ref{Euler-Lagrange equation}) are the first-order differential equations about the evolution coordinate, so only $N$ canonical equations are obtained from Eqs.(\ref{Euler-Lagrange equation}).

\subsection*{Hamiltonian formulation for the NLSE}
After constructing the canonical equations of Hamilton for the first-order differential system, we discuss its application to the NLSE. It is known that the Lagrangian density for the NLSE can be
expressed as~\cite{Anderson-pra-83}
$
l=-\frac{i}{2}(\varphi^*\frac{\partial\varphi}{\partial
t}-\varphi\frac{\partial \varphi^*}{\partial
t})+\frac{1}{2}|\frac{\partial\varphi}{\partial x}|^2-\frac{1}{2}|\varphi|^4
$. The NLSE is complex, and therefore it is an equation with two real functions, the real part of $\varphi$ and its imaginary part. It is convenient to consider
instead the fields $\varphi$ and $\varphi^*$ which are treated as independent from each other. Therefore, $N$  (the components of the quantity) for the NLSE equals two,
i.e., there are two generalized coordinates, $q_1=\varphi^*$ and $q_2=\varphi$, and two generalized momenta can be obtained by using the definition (\ref{conjugate momenta}) as
\begin{equation}\label{generalized momenta for NLSE}
p_1=\frac{i}{2}\varphi,
p_2=-\frac{i}{2}\varphi^*.
\end{equation}
%and two generalized velocities, $\dot{q}_1=\partial\varphi/\partial
%z,\dot{q}_2=\partial\varphi^*/\partial z$.
%Then the generalized momenta can be
%obtained by use of Eq. (\ref{conjugate momenta})
%\begin{equation}\label{generalized momenta for NLSE}
%p_1=\frac{\partial l}{\partial\dot{q}_1}=-\frac{i}{2}\varphi^*,\quad
%p_2=\frac{\partial l}{\partial\dot{q}_2}=\frac{i}{2}\varphi.
%\end{equation}
The Hamiltonian density for the NLSE can be obtained by use of
Eq.(\ref{legendre transform}) as \cite{Hasegawa-book-95}
\begin{equation}\label{Hamiltonian density for NLSE}
h=-\frac{1}{2}\left|\frac{\partial\varphi}{\partial x}\right|^2+
\frac{1}{2}|\varphi|^4.
\end{equation}
%The canonical equations (\ref{canonical equations 1 for 1st order }) and (\ref{canonical equations 2 for 1st order }) will yield two equations, one is the NLSE and the other is its complex conjugate.
If the generalized coordinate $q_1$ and the generalized momentum $p_1$ are taken as the independent variables, the remaining generalized coordinate $q_2$ and the remaining generalized momentum $p_2$ can be expressed via the relations (\ref{generalized momenta for NLSE}) as $q_2=-2ip_1$ and $p_2=-\frac{i}{2}q_1$, respectively. We should also note that the Hamiltonian density $h$ is also the function of $q_{s,x}$, which are independent from $q_s$ and $p_s$. Then for the NLSE, the Hamiltonian density (\ref{Hamiltonian density for NLSE}) should be expressed with the independent variables $q_1,p_1,q_{1,x}$ and $q_{2,x}$ as
\begin{equation}\label{Hamiltonian density for NLSE with independent variables}
h=-\frac{1}{2}q_{1,x}q_{2,x}-2q_1^2p_1^2.
\end{equation}
We should note that $\nu=\mu=1$ and $N=2$ for the NLSE, which means that the equations (\ref{canonical equations 1 for 1st order }) have only one equation, so do Eqs.(\ref{canonical equations 2 for 1st order }). Therefore, the canonical equations (\ref{canonical equations 1 for 1st order }) and (\ref{canonical equations 2 for 1st order }) will yield two equations for the NLSE.
The left side of Eq.(\ref{canonical equations 1 for 1st order })
is obtained as
\begin{equation}
\frac{\delta h}{\delta q_1}=\frac{1}{2}\frac{\partial^2 \varphi}{\partial x^2}+|\varphi|^2\varphi,
\end{equation} and
its right side is
\begin{equation}
-\dot{p}_1+\dot{q}_2\frac{\partial p_2}{\partial q_1}=-i\dot{\varphi}.
\end{equation}
%\begin{eqnarray}
%\frac{\delta h}{\delta \varphi^*}&=&\frac{\partial h}{\partial \varphi^*}-\frac{\partial}{\partial x}\left[\frac{\partial h}{\partial (\partial \varphi^*/\partial x)}\right]-\frac{\partial}{\partial y}\left[\frac{\partial h}{\partial (\partial \varphi^*/\partial y)}\right]\nonumber\\
%&=&-\frac{1}{2}\left(\frac{\partial^2\varphi}{\partial x^2}+\frac{\partial^2\varphi}{\partial y^2}\right)-\left|\varphi\right|^2\varphi.\label{eq1}
%\end{eqnarray}
%\begin{equation}\label{eq2}
%-\frac{\partial{p}_{\varphi^*}}{\partial z}+\frac{\partial\varphi}{\partial z}\frac{\partial p_{\varphi}}{\partial \varphi^*}=i\frac{\partial\varphi}{\partial z}.
%\end{equation}
%Combing Eq.(\ref{eq1}) and Eq.(\ref{eq2}), the NLSE (\ref{NNLSE}) can be derived.
Then the NLSE (\ref{NNLSE}) can be obtained.
%In fact, the complex conjugate of NLSE can also be obtained by using the other canonical equation of Hamilton (\ref{canonical equations 2 for 1st order }). For this purpose, we rewrite the Hamiltonian density (\ref{Hamiltonian density}) in terms of the generalized momenta $p_\varphi,p_{\varphi^*}$ as
%\begin{equation}\label{Hamiltonian density rewritten}
%h=2\left(\left|\frac{\partial p_\varphi}{\partial x}\right|^2+\left|\frac{\partial p_\varphi}{\partial y}\right|^2\right)-8|p_\varphi|^4.
%\end{equation}
Using the other canonical equation, the left side of Eq.(\ref{canonical equations 2 for 1st order }) is obtained as
\begin{equation}\label{derivative of h to varphi}\frac{\delta h}{\delta p_2}=-4q_1^2p_1=-2i|\varphi|^2\varphi^*,\end{equation}
and its right side is
\begin{equation}
\dot{q}_1-\dot{p}_2\frac{\partial q_2}{\partial p_1}+\frac{\partial}{\partial x}\frac{\partial h}{\partial q_{2,x}}\frac{\partial q_2}{\partial p_1}=2\dot{\varphi}^*+i\frac{\partial^2\varphi^*}{\partial x^2}.
\end{equation}
%Inserting Eqs.(\ref{generalized momenta for NLSE}) and (\ref{Hamiltonian density}) into Eq.({\ref{Hamilton canonical equations to NLS2}),
%\begin{eqnarray}
%\frac{\delta h}{\delta p_\varphi^*}&=&\frac{\partial h}{\partial p_\varphi^*}-\frac{\partial}{\partial x}\left[\frac{\partial h}{\partial (\partial p_\varphi^*/\partial x)}\right]-\frac{\partial}{\partial y}\left[\frac{\partial h}{\partial (\partial p_\varphi^*/\partial y)}\right]\nonumber\\
%&=&-2i(\varphi^*\Delta n+\Delta_\bot\varphi^*)\label{eq11}.
%\end{eqnarray}
%The right side of Eq.(\ref{Hamilton canonical equations to NLS2}) is
%\begin{equation}\label{eq22}
%\dot{\varphi}^*-\dot{p}_{\varphi}\frac{\partial \varphi}{\partial p_\varphi^*}=2\dot{\varphi}^*
%\end{equation}
%Combing Eq.(\ref{eq11}) and Eq.(\ref{eq22}),
Then the complex conjugate of the NLSE is also obtained. Therefore, the canonical equations (\ref{canonical equations 1 for 1st order }) and (\ref{canonical equations 2 for 1st order }) are consistent in the sense that the NLSE can be expressed with one of the two canonical equations,
%Eq.(\ref{canonical equations 1 for 1st order })
and its complex conjugate can be expressed with the other.%Eq.(\ref{canonical equations 2 for 1st order }).

In ref.\cite{Hasegawa-book-95}, the canonical equations for the NLSE were considered to be the same as those for the second-order differential system.
Then, according to the definition~\eqref{conjugate momenta}, $p_\varphi=\partial l/\partial\dot{\varphi}$,
%it will be, but . Obviously, this definition is unreasonable.
$p_\varphi$ must be $-i/2\varphi^*$ but not $-i\varphi^*$.
%as can be verified by inserting Eq.(\ref{Lagrangian density}) into Eq.(\ref{conjugate momenta}).
It was artificially doubled in ref.\cite{Hasegawa-book-95} so that $p_\varphi=-i\varphi^*$
in Eq.[5.1.29] (to avoid confusion, we replace the parentheses by the brackets to represent the formulas in the references) to make the NLSE
derived from the canonical equation (\ref{derivative of pi for continuous system}).
In fact, substitution of the Hamiltonian density (\ref{Hamiltonian density for NLSE}) into Eq.(\ref{derivative of pi for continuous system}) only yields $
\frac{i}{2}\frac{\partial \varphi}{\partial
t}+\frac{1}{2}\nabla_\bot^2\varphi+|\varphi|^2\varphi=0,
$
which in fact does not be the NLSE (\ref{NNLSE}). In ref.\cite{Dauxois-book-06}, the canonical equations the authors obtained are Eqs.[3.87] and [3.81], the latter is the same as Eq.(\ref{another form}).
Although the NLSE can be derived from Eq.[3.87], its complex conjugation could not be obtained from the other, Eq.~[3.81]. We now show this claim.
 %that it is impossible to derive the complex conjugate of the NLSE from Eq.(\ref{derivative of q for continuous system}).
 Substituting the Hamiltonian density (\ref{Hamiltonian density for NLSE}) into Eq.(\ref{another form}), the left side of it is obtained as $\frac{\partial \varphi^*}{\partial t}$, and the right side is $\frac{\partial h}{\partial p_{\varphi^*}}=\frac{\partial h}{\partial \varphi}\frac{\partial \varphi}{\partial p_{\varphi^*}}=-2i|\varphi|^2\varphi^*$, where $p_{\varphi^*}=\frac{\partial l}{\partial\dot{\varphi}^*}$. Then the equation
$
-\frac{i}{2}\frac{\partial \varphi^*}{\partial
t}+\left|\varphi\right|^2\varphi^*=0
$
can be obtained,
which is absolutely not the complex conjugate of the NLSE. Therefore, the canonical equations for the NLSE obtained in ref.\cite{Dauxois-book-06} are inconsistent.

\subsection*{Conclusion}
Summarizing, we obtain a new form of canonical equations of Hamilton with the complete symmetry, which are valid not only
for the first-order differential system, but also for the second-order differential
system. The conventional form of the canonical equations without the symmetry are only for the second-order differential
system. The number of the canonical equations for the first-order differential
system is $N$ rather than
$2N$ like the case for the second-order differential
system. The canonical equations for the NLSE are two equations, which are consistent in the sense that the NLSE can be expressed with one of the canonical equations and its complex conjugate can be expressed with the other. The Hamiltonian formulation for the continuous system can also be extended to the discrete system.

 It is well known that the symmetry plays an important role in theoretical physics~\cite{Gross-colloquium-paper-1996}. The search for and the discovery of new symmetries promote the exploration of fundamental laws of physics. Based on the idea, the conical equations of Hamilton with the complete symmetry found by us might find their appropriate position in modern theoretical physics.
\subsection*{Acknowledgements}
This research was supported by the National Natural Science Foundation of China (Grant Nos. 11074080 and 11274125).

\end{document}